\documentclass[review]{elsarticle}
\journal{}
\biboptions{authoryear}

\usepackage[margin=2.5cm]{geometry}
\usepackage{comment}
\usepackage[numbered]{bookmark}
\usepackage{amsmath,amssymb,amsthm}
\usepackage{soul}
\usepackage[ruled,vlined]{algorithm2e}

\usepackage{xfrac}
\usepackage{color, colortbl}
\usepackage[makeroom]{cancel}
\usepackage{bbding}
\usepackage{rotating}
\usepackage{subfigure}
\usepackage{multirow}
\usepackage{array}
\usepackage{graphicx}
\usepackage{animate}
\usepackage{tikz}
\usepackage{pgfplots}
\usepackage{mathtools}
\usetikzlibrary{fit,shapes.misc}
\usepackage{pdflscape}
\usepackage[utf8x]{inputenc}
\usepackage{lmodern,textcomp}

\renewcommand\appendix{\par
  \setcounter{section}{0}
  \setcounter{subsection}{0}
  \setcounter{figure}{0}
  \setcounter{table}{0}
  \renewcommand\thesection{Appendix \Alph{section}}
  \renewcommand\thefigure{\Alph{section}\arabic{figure}}
  \renewcommand\thetable{\Alph{section}\arabic{table}}
}
\theoremstyle{plain}
\theoremstyle{plain}
\theoremstyle{plain}

\usepackage{hyperref}
\hypersetup{
  colorlinks   = true, 
  urlcolor     = blue, 
  linkcolor    = blue, 
  citecolor   = blue!70 
}

\begin{document}
\begin{frontmatter}

\title{A new framework of sensor selection for developing a fault detection system based on data-envelopment analysis}

\author[Amiraddress]{Amir Eshaghi Chaleshtori\href{https://orcid.org/0000-0002-5495-5748}{\includegraphics[scale=0.6]{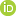}}}
\ead{eshaghi-ch@email.kntu.ac.ir}
\author[Amiraddress]{Abdollah Aghaie\href{https://orcid.org/0000-0002-1022-7757}{\includegraphics[scale=0.6]{Figures/orcid.png}}\corref{correspondingauthor}}
\ead{aaghaie@kntu.ac.ir}
\cortext[correspondingauthor]{Corresponding author}

\address[Amiraddress]{School of Industrial Engineering,K.N.Toosi University of Technology, Tehran, Iran}

\begin{abstract}
Several methods have been proposed to identify which sensor sets are optimal for finding and localizing faults under different conditions for mechanical equipment. In order to preserve acceptable performance while minimizing costs, it is crucial to identify the most effective set of sensors available. Nevertheless, some sensor sets can behave differently in fault detection because of uncertainty in the measurement data contaminated by noise interference. This paper develops new sensor selection models using Data Envelopment Analysis (DEA), which has proven helpful as a management approach for determining an optimal number of sensors, associated costs, and sensor performance in the fault diagnosis. We propose four linear optimization models for sensor selection to design the fault detection system. The validity of the presented models is demonstrated using a vibration dataset collected from a gearbox. Based on the case study results, the proposed methods are effectively superior to the other comparison sensor selection methods in fault detection accuracy.	
\end{abstract}
\end{frontmatter}

\section{Introduction}\label{sec:Intr}
Failure detection of complex systems using the signals collected by multiple sensors is essential in identifying incipient faults to prevent catastrophic malfunctions, reducing productivity losses and interruptions to operations while ensuring human safety\citep{buchaiah2022bearing}. Adding more sensors may provide accurate fault detection results but will also lead to higher sensor costs and occupy more space \citep{bhushan2008robust,jung2020sensor}. Consequently, selecting the most appropriate sensors is critical to satisfying the failure analysis objectives while minimizing the related costs.
The sensor selection problem takes various forms, ranging from minor to larger systems\citep{kulkarni2021sensor}. In small systems, the selection of sensors is primarily experimental and relatively straightforward, while for large complex systems, it is more challenging and necessitates further analysis\citep{kulkarni2021sensor}. However, most of the suggested sensor selection methods are not scalable, i.e., the selection method for small systems is not suitable for large complex systems and vice versa\citep{clark2020sensor,kulkarni2021sensor,liesegang2021sensor}.
Moreover, previously sensor selection methods were heavily dependent on the decision maker's expertise, resulting in subjective decisions\citep{chauhan2020mobile,ye2020complexity,kulkarni2021sensor}. However, an essential consideration when dealing with the sensor selection problem is the adverse effect of uncertainty in the collected data and measurement noise on diagnostic accuracy \citep{yamada2021fast,nagata2022data}. Significant uncertainties impede the identification of minor faults\citep{yamada2021fast}. Although a set of sensors satisfies fault detection requirements, it is not guaranteed that diagnosis systems utilizing these sensors will perform as expected when the uncertainties in the collected data, such as measurement noise, are considered. Hence, it is necessary to consider more realistic parameters in the sensor selection process than simply selecting a minimum number of sensors that provide fault detection under ideal conditions\citep{yamada2021fast}.
Moreover, Bhushan et al. \citep{bhushan2008robust} emphasized that the cost of a system is always an essential consideration in its design. The same is true of fault detection systems as well. Thus, it is imperative to regard the associated sensor costs\citep{kathiroli2021energy}. By translating these requirements into the sensor selection problem, we would ensure that the fault diagnostic system could be implemented with acceptable performance\citep{jung2020sensor}. This paper utilizes Data Envelopment Analysis (DEA) models to develop sensor selection methods considering the properties of the sensors and related costs. A significant benefit of DEA for the sensor selection is that it will provide a composite measure of total performance, enabling a straightforward comparison among sensors\citep{fancello2020data,yamada2021fast}. DEA also provides information on the sensor performance indicators included in the final score\citep{easton2002purchasing}. This information allows the expert to decide based on specific measures\citep{horvathova2020comparison}. Moreover, DEA provides metrics of efficiency (output/inputs) so that both input and output variables are taken into account to make an accurate comparison between sensors\citep{de2013negative}. In summary, the current work makes the following contributions:

\begin{itemize}
\item This paper establishes a data-driven method that considers the properties of sensor data and the associated costs.
\item This paper develops DEA models for selecting the informative sensors to design a fault diagnosis system.
\item Outcomes from experiments on a multi-sensor dataset reveal that the presented methods are highly effective at diagnosing faults.
\end{itemize}
This study is divided into the following sections. Section \ref{sec:LitRev} surveys the literature on the sensor selection problem. Section \ref{sec:ProposedMethod} introduces the fundamental theories of the proposed approaches. Section \ref{sec:Experiments} describes the details related to the proposed sensor selection methods. Section \ref{sec:ResultsAndDiscuss} discusses the experimental validation of the suggested methods. Lastly, the article outlines its conclusions in section \ref{sec:Conclusion}.
\section{Literature review}\label{sec:LitRev}
Sensor selection methods are examined in many fields for choosing informative sensors, such as prognostic health management\citep{cheng2010sensor,guan2020comprehensive,chaleshtori2022data}, boost-stage rocket engines\citep{sowers2012systematic}, turbo-fan engines \citep{sowers2009expanded},  aircraft engine health estimation \citep{simon2009systematic}, plant equipment \citep{riedel2015knowledge}, vehicle design \citep{santi2005optimal}, and the polymer electrolyte membrane (PEM) fuel cell \citep{mao2016selection,mao2018effectiveness}. Moreover, Graph-based methods\citep{shieh2001selection,zhang2007methodology}, semantic approaches\citep{schmidt2001build,kertiou2018dynamic}, greedy algorithms\citep{zhang2017sensor,clark2020multi,jung2020sensor},knowledge-based methods\citep{riedel2015knowledge,jones2018straightforward}, and grey clustering methods\citep{guan2020comprehensive} are the most widespread sensor selection methods in literature.Furthermore, many optimization methods are employed based on convex relaxation methods to solve sensor selection problems (as a heuristic alternative to brute-force search)\citep{joshi2008sensor,nonomura2021randomized}, greedy methods \citep{manohar2018data,saito2021determinant} and proximal optimization algorithms \citep{dhingra2014admm,nagata2021data}. However, various objective metrics concerning sensor selection algorithms exist. The Fisher information matrix, ordinary linear least square estimation \citep{peherstorfer2020stability,nakai2021effect}, and a steady state error covariance matrix of the Kalman filter \citep{ye2018complexity} exemplify sensor selection objectives. Despite these established methodologies, noise interference could only be considered occasionally in formulations. However, there can be some challenges associated with noise due to the differences between models and phenomena when processing acoustic signals \citep{o2016distributed}, vibrations \citep{castro2013robustness}, and data-driven modelling \citep{saito2021determinant}. Liu et al.\citep{liu2016sensor} considered an assumption of noise with low correlation, and Joshi et al.\citep{joshi2008sensor} developed algorithms based on a general kernel of noise covariance between sensors.\\ The purpose of this paper is to develop sensor selection methods to take into account the properties of the sensors and associated costs using Data Envelopment Analysis (DEA) models. The fault diagnostic system could be implemented with acceptable performance if we translated these requirements into the sensor selection problem. The current work will be described in the following.
\section{The suggested sensor selection method}\label{sec:ProposedMethod}
\begin{figure}[ht!]
\begin{center}
\includegraphics[width=0.95\textwidth,trim={0.5cm 4.5cm 0.7cm 2cm},clip]{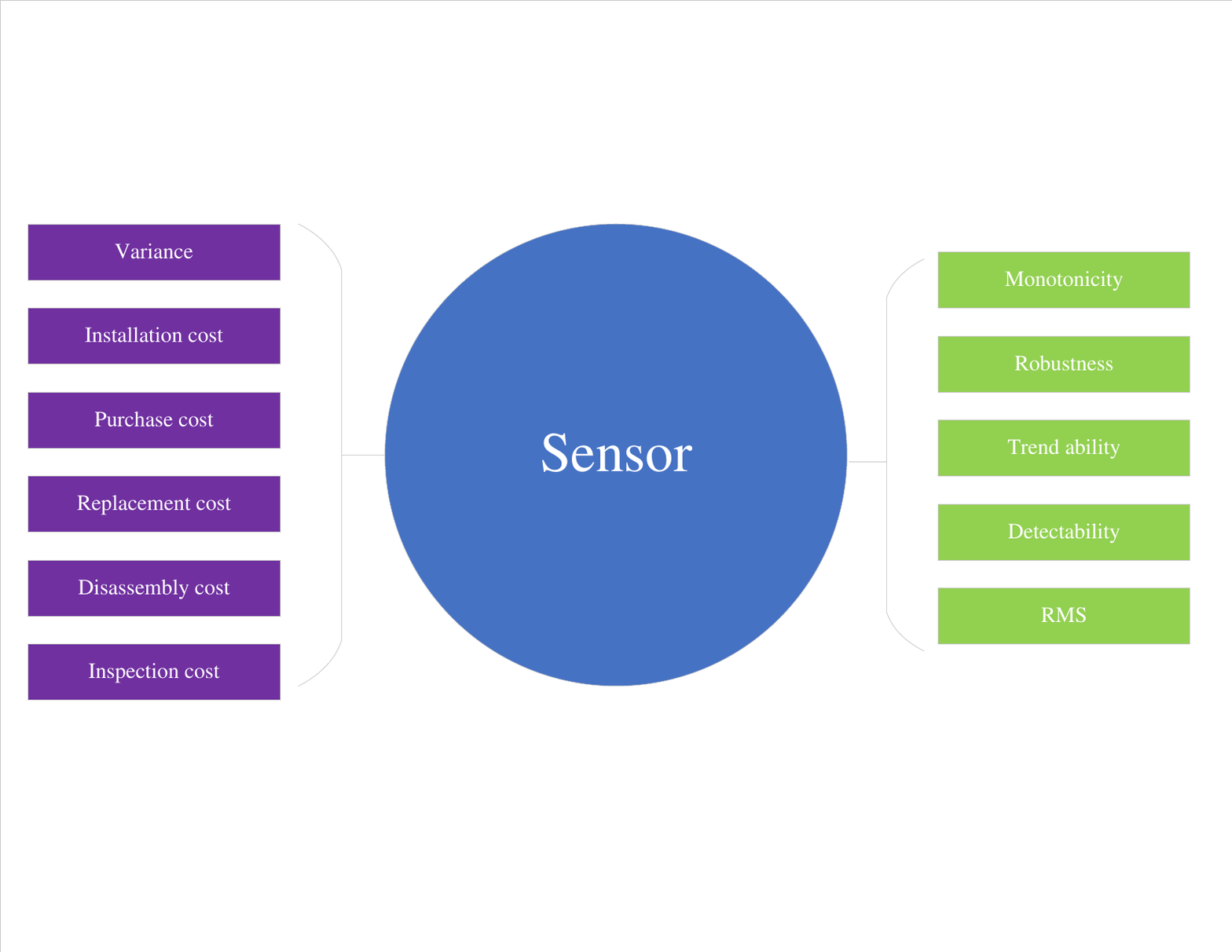}
\caption{The proposed decision making unit(Sensor).}
\label{Fig:FlowchartDMU}
\end{center}
\end{figure}

\begin{figure}[ht!]
\begin{center}
\includegraphics[width=0.95\textwidth,trim={0.5cm 0.8cm 0.7cm 0.7cm},clip]{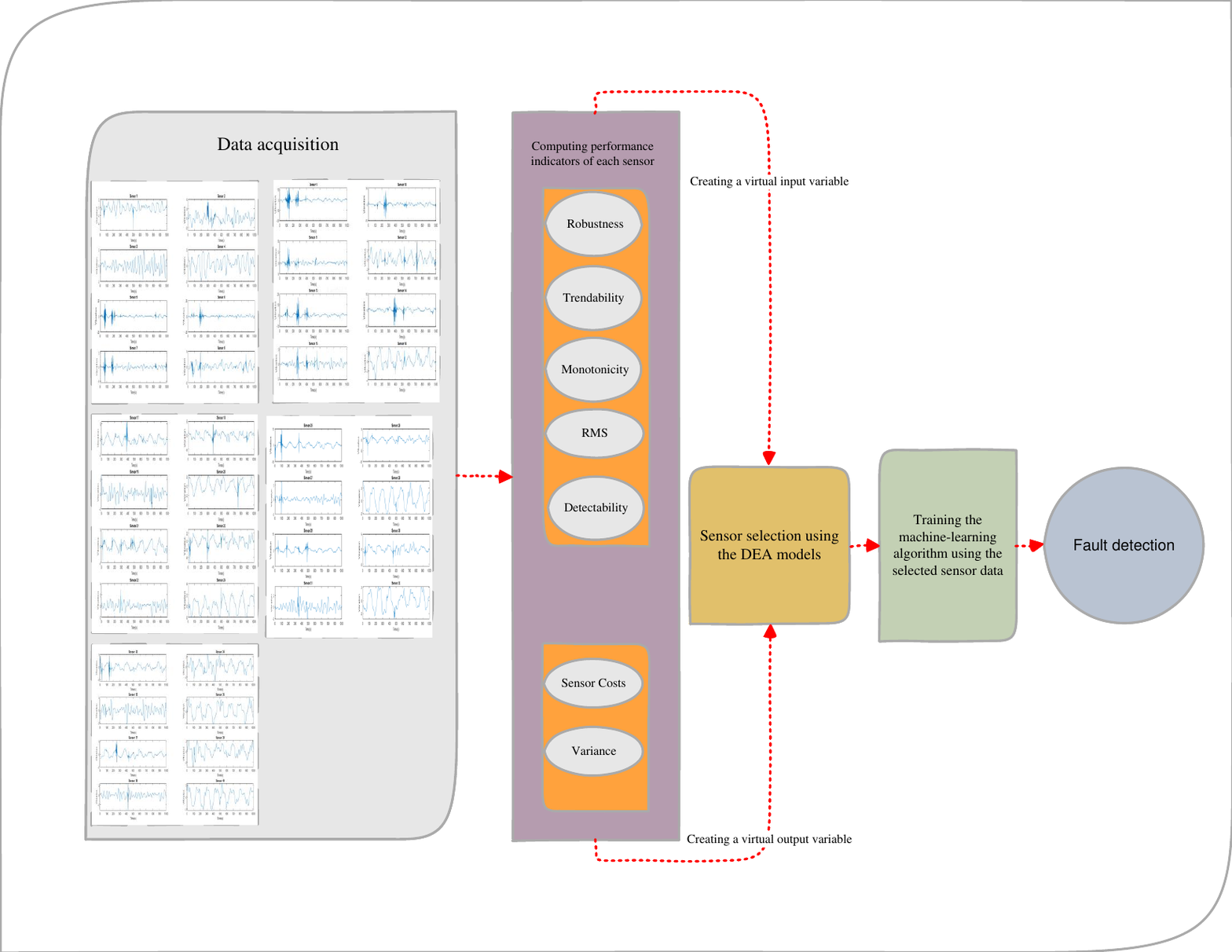}
\caption{Flowchart of the proposed methodology.}
\label{Fig:Flowchart}
\end{center}
\end{figure}

This section explains the suggested framework and how to detect faults using supervised machine learning algorithms. The framework involves two stages. In the first stage, DEA models select significant sensors. It incorporates all the characteristics of sensors and allows a final decision based on the system state.
The following parameters are considered when designing a practical sensor selection framework: monotonicity, robustness, trendability, detectability, variance, root mean square and sensor costs.
In detail, these parameters are calculated for each sensor. Then for each sensor, a virtual input variable, defined as the weighted sum of the monotonicity, robustness, trendability, detectability, and the root mean square, is created. Similarly, a virtual output variable, defined as the weighted sum of sensor costs and the variance, is created for each sensor. Figure \ref{Fig:FlowchartDMU} shows the defined virtual variables for each sensor. Then, the virtual input and output variables are fed into DEA models to find the efficient sensors according to the system state and sensor requirements for condition monitoring. In the second stage, machine-learning algorithms are trained using the most efficient sensor data to determine equipment failures and test the accuracy of sensor selection models. Figure \ref{Fig:Flowchart} shows a schematic illustration of the proposed method. The considered parameters are defined in the following subsection.

\subsection{The performance parameters of sensor selection framework}\label{subsec:PerformanceCriteria}
\begin{itemize}
\item Monotonicity:In order to extract information from trends that are increasing or decreasing, the monotonicity metric is used.It is expressed as follows:
\begin{equation}\label{Eq:Monotonicity}
\text { Monotonicity }=\frac{1}{k} \sum_{j=1}^k \sum_{x \in c_j} \frac{1}{N-1}\left|\# \frac{d}{d x}>*-\# \frac{d}{d x}<\cdot\right|
\end{equation}
where $k$ and $N$ represents the number of states and the number of data within each state,$c_{j}$ represents the $j^{th}$ state, and $\frac{d}{d x}=x_{n+1}-x_{n}$ is the difference between two successive data. The value of the monotonicity is between 0 to 1.
\end{itemize}

\begin{itemize}
\item Robustness:In order to calculate the robustness, the trend and residual components are extracted. The residual component ($res_{x}$) of the signal $x$ is extracted by subtracting the smoothed signal $sm_{x}$(trend) from the original signal $x$, as follows. This paper utilizes the wavelet denoising algorithm proposed by Ref.\citep{bnou2020wavelet} to estimate the smoothed signal.
\begin{equation}\label{Eq:Robustness}
\text { Robustness }=\frac{1}{k} \sum_{j=1}^k \frac{1}{N} \sum_{x \in c_j} e^{-\left|\frac{r e s_x}{x}\right|} \quad ; \text { res }_x=x-s m_x
\end{equation}
where $k$ and $N$ represents the number of states and the number of data within each state,$c_{j}$ represents the $j^{th}$ state.
\end{itemize}

\begin{itemize}
\item Trendability:According to the trendability metric, a given signal and its lagged version exhibit similarity over successive time intervals. The signal is measured in terms of the relationship between its current and past values.
\begin{equation}\label{Eq:Trendability}
\text { Trendibility }=\frac{1}{k} \sum_{j=1}^k \sum_{t=\cdot}^T \sum_{\tau=\cdot}^{T-1} \sum_{x_t \in c_j}\left|x_t\left(x_{t-\tau}\right)^T\right|
\end{equation}
where $k$ and $N$ represents the number of states and the number of data within each state,$c_{j}$ represents the $j^{th}$ state.$T$ represents the time length of the signal and $\tau$ shows the time lag.
\end{itemize}

\begin{itemize}
\item Detectability:It refers to the ability to distinguish between different states. The Fisher discriminant ratio is used to compare the scatter between states and the scatter within states \citep{powell2013user}. This ratio forms the basis of the detectability metric. As a result, we calculate it as follows:
\begin{equation}\label{Eq:Distinguishability}
\text { Detectability }=\frac{\sum_j n_j\left(\bar{x}_j-\overline{\bar{x}}\right)\left(\bar{x}_j-\overline{\bar{x}}\right)^T}{\sum_j \sum_{x \in c_j}\left(x-\bar{x}_j\right)\left(x-\bar{x}_j\right)^T}
\end{equation}
where $\bar{x}_{j}$ indicate the average value of $j^{th}$ state,$\overline{\bar{x}}$ represents the mean value of all data,and $n_{j}$ is the number of data in the $j^{th}$ state.
\end{itemize}

\begin{itemize}
\item Variance:The variance is a measure of dispersion, which indicates how far the signal deviates from its average.
\begin{equation}\label{Eq:var}
\text { Variance }=\frac{1}{k} \sum_{j=1}^k \frac{1}{N-1} \sum_{x \in c_j}\left(x-\bar{x}_j\right)\left(x-\bar{x}_j\right)^T
\end{equation}
where $k$ and $N$ represents the number of states and the number of data within each state,$c_{j}$ represents the $j^{th}$ state, and $\bar{x}_{j}$ indicate the average value of $j^{th}$ state.
\end{itemize}

\begin{itemize}
\item Root mean square(RMS):This term refers to the calculated overall energy and amplitude of a given signal. According to the RMS, the following formula is provided:
\begin{equation}\label{Eq:RMS}
R M S=\frac{1}{k} \sum_{j=1}^k \sqrt{\frac{1}{N} \sum_{x \in c_j} x^{2}}
\end{equation}
\end{itemize}

\begin{itemize}
\item Sensor costs:The cost of any system is always a factor in its design; the same applies to fault detection systems. Sensor costs are typically evaluated by considering the costs associated with sensor acquisition, installation, communication, replacement, disassembly, and inspection\citep{xu2005optimal,zhang2018survey}.
\end{itemize}
For the bearing fault diagnosis, higher values of monotonicity,trendability, robustness, detectability, and RMS are desirable\citep{duong2018reliable}. Moreover, lower variance and sensor cost values are preferred \citep{kulkarni2021sensor}. Thus, the following DEA models have output variables: monotonicity, trendability, robustness, detectability, and RMS, while input variables are variance and sensor costs.

\subsection{The DEA models}\label{subsec:DEA}
The DEA approach uses linear programming to convert multiple input and output variables into a quantitative measure of productivity efficiency\citep{easton2002purchasing}. Efficiencies are used to measure how efficiently resources are utilized and how processes are performed when they transform input variables into output variables. Efficiency is a relative concept since it implies that a unit's performance is compared against a standard. The DEA analyzes the relative efficiency of decision-making units (DMUs). We consider sensors as DMUs in our proposed application, shown in Figure \ref{Fig:FlowchartDMU}.

\subsubsection{The CCR model}
Constant Returns to Scale, or CCR, is the first basic DEA model developed by Charnes et al.\citep {charnes1978measuring}. In a typical DEA model, input and output variables are predetermined, and a comparison of the efficiency of similar elements is attempted. Thus, DMUs are considered the elements to be compared. Each DMU's efficiency is defined as the weighted sum of output variables divided by the weighted sum of inputs. An efficient unit has a score of one. As an additional benefit of DEA, it does not require any subjective weighting procedures, and an overall performance score is derived as an efficiency score for a DMU. By assigning weights to input and output variables using the calculation model, the decision-maker does not have to make subjective decisions to maximize the DMU's efficiency. The traditional DEA model utilizes a linear divisive programming approach to calculate the relative efficiency of any DMU.
\begin{equation}\label{Eq:CCR}
\begin{gathered}
\max z_l: \frac{\sum_r u_r y_{r l}}{\sum_s v_s o_{s l}} \\
\text { Subject to: } \\
0\leq \frac{\sum_r u_r y_{r l}}{\sum_s v_s o_{s l}} \leq 1 \quad \forall l=1, \ldots, k \\
u_r, v_s \geq 0 \quad \forall r=1,2, \ldots, N, \forall s=1,2, \ldots, M
\end{gathered}
\end{equation}
In the objective function of the CCR model, the efficiency of the $l^{th}$ DMU is maximized. Using the Equation(\ref{Eq:CCR}), all DMUs are defined based on the weights selected for $l$. Equation (\ref{Eq:CCR}) provides that the efficiency of all DMUs is between 0 and 1. In the final constraint, the weights of output and input variables must all be positive.
\subsubsection{The BCC model}
Banker et al.\citep{banker1984some} developed the BCC model to calculate efficiencies that did not result in a proportionate change in input or output variables according to changes in input or output variables. Depending on the approach, input or output variables may be the focus of the BCC model. The input-oriented model produces the same inputs with the minimum amount that can be formulated as follows:
\begin{equation}\label{Eq:IO-BCC}
\begin{gathered}
\max z_l: \frac{\sum_r u_r y_{r l}+u_0}{\sum_s v_s o_{s l}} \\
\text { Subject to: } \\
0\leq \frac{\sum_r u_r y_{r l}+u_0}{\sum_s v_s o_{s l}} \leq 1 \quad \forall l=1, \ldots, k \\
u_r, v_s \geq 0 \quad \forall r=1,2, \ldots, N, \forall s=1,2, \ldots, M \\
u_0 \text { : free }
\end{gathered}
\end{equation}
while the output-oriented model is designed to maximize output variables with the least amount of inputs which can be formulated as follows:
\begin{equation}\label{Eq:OO-BCC}
\begin{gathered}
\max z_l: \frac{\sum_r u_r y_{r l}}{\sum_s v_s o_{s l}+v_0} \\
\text { Subject to: } \\
0\leq \frac{\sum_r u_r y_{r l}}{\sum_s v_s o_{s l}+v_0} \leq 1 \quad \forall l=1, \ldots, k \\
u_r, v_s \geq 0 \quad \forall r=1,2, \ldots, N, \forall s=1,2, \ldots, M \\
v_0: \text { free }
\end{gathered}
\end{equation}
Both objective functions have a maximum efficiency of the $l^{th}$ DMU. Using Equations (\ref{Eq:IO-BCC}) and (\ref{Eq:OO-BCC}), each DMU is defined in terms of the weights selected for $l$. Accordingly, the efficiency of all DMUs lies between 0 and 1. In both models, the final constraints require all output and input variables to have a positive weight.
\subsubsection{The additive model}
Additive DEA aims to maximize output variables for a minimum amount of input variables\citep{jin20232,chen2023regret}. The objective function of this model is a summation of weighted input and output variables that can be expressed as follows:
\begin{equation}\label{Eq:Additive}
\begin{gathered}
\max z_l: \sum_r u_r y_{r k}-\sum_s v_s o_{s l}-w_0 \\
\text { Subject to: } \\
\sum_r u_r y_{r l} \leq \sum_s v_s o_{s l}+w_0 \quad \forall l=1, \ldots, k \\
u_r, v_s \geq 1 \quad \forall r=1,2, \ldots, N, \forall s=1,2, \ldots, M \\
w_0 \text { : free }
\end{gathered}
\end{equation}
According to the objective function, the efficiency of the $l^{th}$ DMU is maximized. The DMUs are a function of the weights selected for $l$, using the Equation(\ref{Eq:Additive}). A reference Equation (\ref{Eq:Additive}) states that the efficiency of all DMUs is between 0 and 1. All output and input variables must have positive weights in the final constraint.
\section{Experiments}\label{sec:Experiments}
\subsection{Dataset for experiments}\label{subsec:ExperimentDataSet}
\begin{figure}[ht!]

    \begin{center}
    \subfigure[Sensor1]{
    \includegraphics[trim={1.2cm 1cm 0cm 0cm},clip,width=0.5\columnwidth]{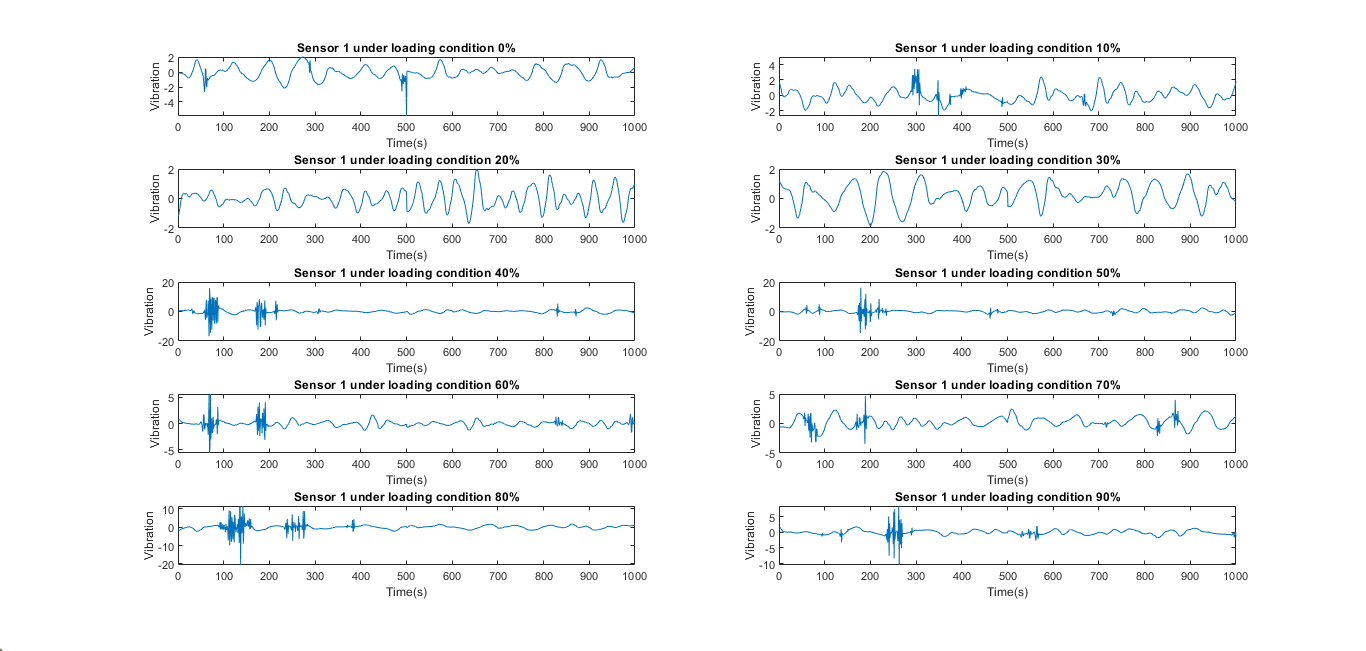}
    \label{Fig:Sensor1}
    }
    \subfigure[Sensor2]{
    \includegraphics[trim={1.2cm 1cm 0cm 0cm},clip,width=0.5\columnwidth]{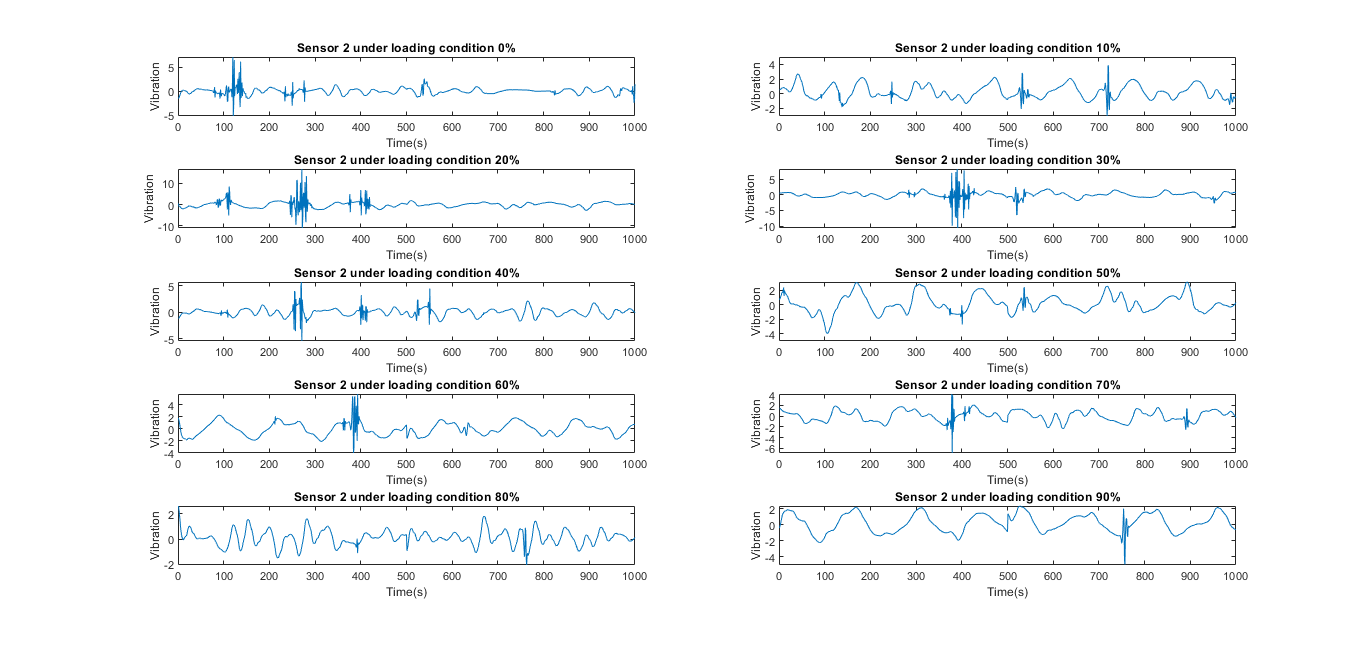}
    \label{Fig:Sensor2}
    }
    \subfigure[Sensor3]{
    \includegraphics[trim={1.2cm 1cm 0cm 0cm},clip,width=0.5\columnwidth]{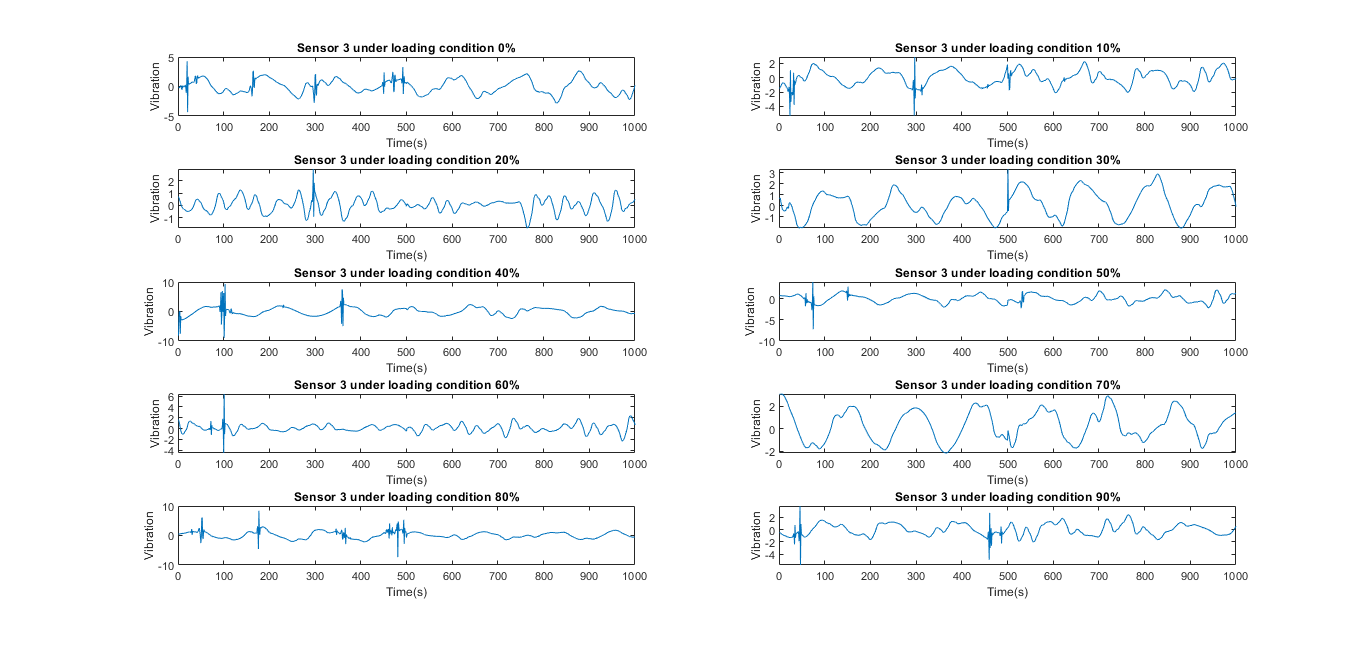}
    \label{Fig:Sensor3}
    }
    \subfigure[Sensor4]{
    \includegraphics[trim={1.2cm 1cm 0cm 0cm},clip,width=0.5\columnwidth]{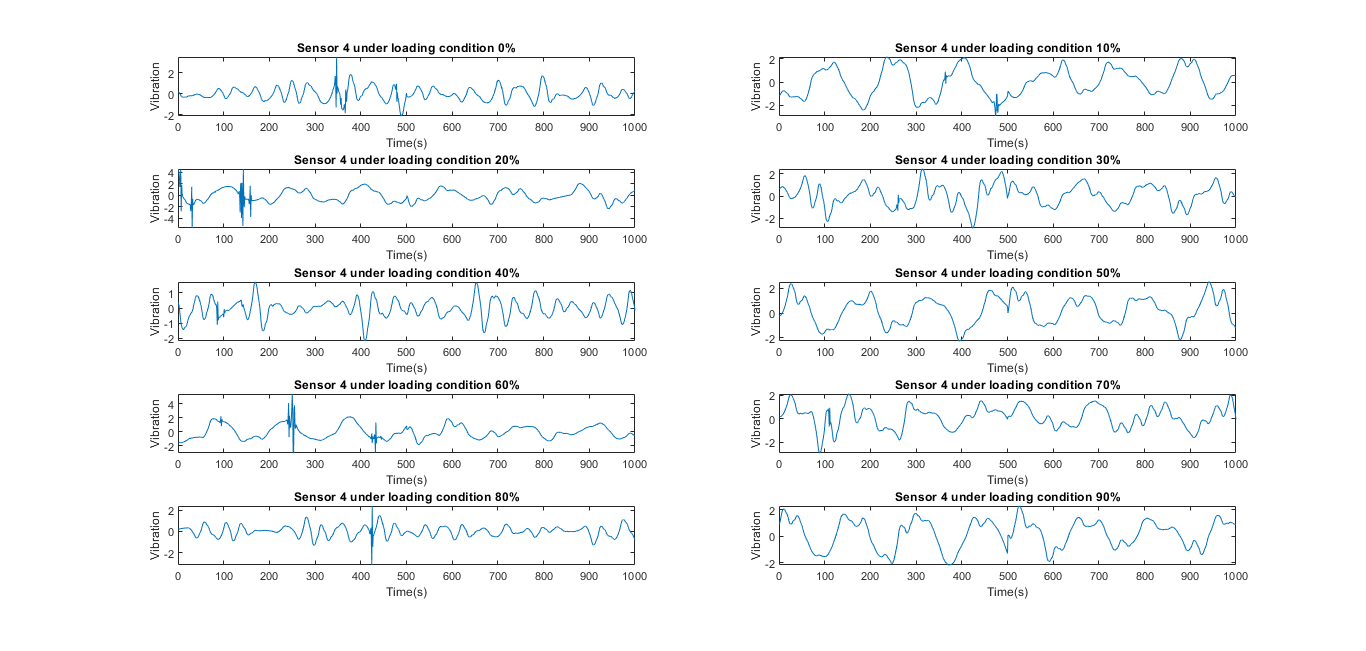}
    \label{Fig:Sensor4}
    }
    \caption{The collected vibration data  under varying loading conditions.}
    \label{Fig:RawData}
    \end{center}
\end{figure}

The experiments are conducted using Spectra Quest's Gear Fault Diagnosis Simulator to create gear fault diagnostic data for condition monitoring.
The Gearbox Fault Diagnostic Simulator (GFDS) developed by Spectra Quest has been designed to simulate industrial gearboxes for prognostics and diagnostics. Data are generated using GFDS for two cases: a healthy condition and a condition with a broken tooth, as shown in Table\ref{tab:StateLabel} . Four vibration sensors are placed on the body of the gearbox in four different directions, and the vibration signals are recorded under varying loading conditions from zero loads to 90\%\citep{malik2019feature}. Hence, 40 vibration data have been derived from the sensors under various loading conditions. In our study, we select 500 consecutive sets of data points for each gearbox condition to examine the performance of the proposed sensor selection method. Figures \ref{Fig:RawData} show the vibration data collected under varying loading conditions from zero loads to 90\%.
\begin{table}[ht!]
    \centering
     \caption{An overview of each bearing state and its corresponding label in the dataset.}
    \label{tab:StateLabel}
    \begin{tabular}{cc}
    \hline
        Status &label\\
        \hline
        Healthy condition  &1\\
        Broken tooth condition   &2\\
       \hline
    \end{tabular}
\end{table}

\subsection{Overview of the evaluation criteria}\label{susec:EvaluationCriteria}
There have been several performance metrics proposed based on the binary confusion matrix. Five of these metrics are the focus of our study, which can be summarized as follows:
\begin{itemize}
\item Accuracy:Generally, the accuracy metric is defined as the proportion of correct predictions to the number of observations that can be defined as follows:
\begin{equation}\label{Eq:Accuracy}
\text {Accuracy}=\frac{TP+TN}{TP+TN+FP+FN}
\end{equation}
\end{itemize}

\begin{itemize}
\item Recall:The recall is used to determine the percentage of correctly classified patterns that can be expressed as follows:
\begin{equation}\label{Eq:RecallPlus}
Recall^{+}=\frac{TP}{TP+FN}
\end{equation}
\begin{equation}\label{Eq:RecallMinus}
Recall^{-}=\frac{TN}{TN+FP}
\end{equation}
\end{itemize}

\begin{itemize}
\item Precision:In classification problems, \textit{precision} is a proportion of positive patterns correctly predicted based on  total predicted patterns that can be defined as follows:
\begin{equation}\label{Eq:PrecisionPlus}
Precision^{+}=\frac{TP}{TP+FP}
\end{equation}

\begin{equation}\label{Eq:PrecisionMinus}
Precision^{-}=\frac{TN}{TN+FN}
\end{equation}
\end{itemize}

\begin{itemize}
\item F-score:In this metric, recall and precision values are expressed as a harmonic mean that can be expressed as follows:
\begin{equation}\label{Eq:FscorePlus}
F^{+}=2 . \frac{precision^{+}. recall^{+}}{precision^{+} + recall^{+}}
\end{equation}
\begin{equation}\label{Eq:FscoreMinus}
F^{-}=2 . \frac{precision^{-}. recall^{-}}{precision^{-} + recall^{-}}
\end{equation}
\end{itemize}

\begin{itemize}
\item The area under the ROC curve(AUC):The AUC is an index used to measure the performance of a statistical diagnosis algorithm. One can determine how effectively a particular classifier can quantitatively separate two groups based on ROC curves. The ROC curve shows the trade-off between the probability of detection or the true positive rate, also known as sensitivity or recall, and the probability of a false alarm or false positive rate.
\end{itemize}

\subsection{Procedures for experiments}\label{subsec:ExProcedure}
Sensor selection begins with an analysis of the system and determines the sensor requirements for condition monitoring that allows for identifying the most informative sensor by DEA models. For this study, we simulate the sensor costs reported in Table\ref{tab:SimulatedCosts}. Additionally, we calculate a number of parameters related to the gearbox vibration data, including monotonicity, robustness, trendability, detectability, variance, and RMS.

\begin{table}[ht!]
\tiny{
    \centering
     \caption{The simulated sensor costs(\$).}
    \label{tab:SimulatedCosts}
    \begin{center}
    \begin{tabular}{ccccccc}
    \hline
         Sensor&Load condition(\%)&Purchase cost&Installation cost&Replacement cost&Disassembly cost&Inspection cost\\
        \hline
        \multirow{10}{*}{1}&0&201.48&67.85&82&186.86&168.27\\
        ~&                  10&199.59&91.54&338.57&133.39&131.87\\
        ~&                  20&212.26&79.26&51.39&113.11&191.19\\
        ~&                  30&213.24&77.49&282.47&135.34&153\\
        ~&                  40&193.17&95.86&295.19&93.21&128.05\\
        ~&                  50&203.16&64.29&310.61&98.08&208.64\\
        ~&                  60&201.70&87.86&75.33&209.06&200\\
        ~&                  70&208.33&87.69&169.93&210.99&155.37\\
        ~&                  80&210.41&69.02&127.96&158.80&165.28\\
        ~&                  90&211.90&78.39&290.02&88.19&160.43\\
       \hline
       \multirow{10}{*}{2}&0&196.11&	53.79&	179.42&	112.16&	108.46\\
        ~&                  10&209.43&	52.70&	323.19&	128.38&	121.27\\
        ~&                  20&208.62&	76.54&	104.55&	192.50&	144.52\\
        ~&                  30&192.37&	88.96&	129.14&	82.11&	111.58\\
        ~&                  40&190.93&	96.70&	93.66&	85.89&	195.67\\
        ~&                  50&203.45&	56.50&	90.82&	103.15&	106.68\\
        ~&                  60&218.67&	78.44&	310.79&	168.93&	110.95\\
        ~&                  70&198.23&	73.47&	223.91&	180.25&	103.39\\
        ~&                  80&206.31&	50.60&	214.96&	168.74&	111.19\\
        ~&                  90&194.39&	66.86&	93.49&	141.78&	139.69\\
       \hline
        \multirow{10}{*}{3}&0&211.79&	58.11&	305.91&	154.94&	122.62\\
        ~&                  10&195.42&	89.71&	236.62&	120.60&	206.50\\
        ~&                  20&203.70&	65.56&	155.29&	182.02&	138.94\\
        ~&                  30&210.07&	76.43&	203.97&	105.89&	105.32\\
        ~&                  40&216.40&	58.28&	170.54&	174.09&	203.97\\
        ~&                  50&218.66&	80.10&	72.79&	105.14&	214.23\\
        ~&                  60&205.06&	63.15&	121.97&	130.48&	140.13\\
        ~&                  70&191.57&	82.70&	87	  & 165.71&	95.22\\
        ~&                  80&191.93&	84.46&	105.17&	186.89&	115.35\\
        ~&                  90&195.50&	87.41&	121.99&	91.11 &	135.99\\
       \hline
        \multirow{10}{*}{4}&0&214.74&	72.53&	175.18&	207.33&	161.50\\
        ~&                  10&195.39&	54.19&	64.90 &	186.27&	115.92\\
        ~&                  20&213.87&	61.45&	320.81&	146.69&	162.59\\
        ~&                  30&195.04&	95.67&	333.44&	139.71&	177.44\\
        ~&                  40&217.67&	57.62&	197.26&	141.21&	110.38\\
        ~&                  50&198.55&	91.29&	196.78&	121.97&	96.09\\
        ~&                  60&193.49&	76.92&	151.32&	149.67&	120.64\\
        ~&                  70&195.29&	99.81&	320.02&	149.98&	123.67\\
        ~&                  80&207.33&	53.91&	160.77&	192.01&	138.11\\
        ~&                  90&202.62&	72.13&	83.36 &	188.89&	149.58\\
       \hline
    \end{tabular}
    \end{center}
    }
\end{table}
Comparative analysis is performed among the proposed sensor selection models(The CCR, The IO-BCC, the OO-BCC, and the additive) and traditional methods, such as sensor selection based on correlation analysis implemented by Khalid et al.\citep{khalid2021real}, the mRMR algorithm proposed by Peng et al.\citep{peng2005feature}, and the Extra Tree Classifier technique(ETC) suggested by Sharaff\citep{sharaff2019extra}. Moreover, we apply the well-known supervised machine learning classifiers for the fault classification, including the support vector machine (SVM), the k-nearest neighbor classifier (KNN), and the naive Bayes algorithm. It is crucial to tune the hyperparameters of machine learning classifiers. As a result, we use a five-fold cross-validation method to optimize the hyperparameters to minimize the adverse effects of data volatility. Specifically, we utilize four folds for training and the left for testing. Once the parameters have been evaluated five times for all training data, we will select the parameters that produce the best results for the classification of the test data.
\section{Results and discussions}\label{sec:ResultsAndDiscuss}
\subsection{Results}\label{subsec:Results}
\begin{figure}[ht!]
    \begin{center}
    \subfigure[CCR-KNN]{
    \includegraphics[trim={1.2cm 1cm 2cm 2cm},clip,width=0.3\columnwidth]{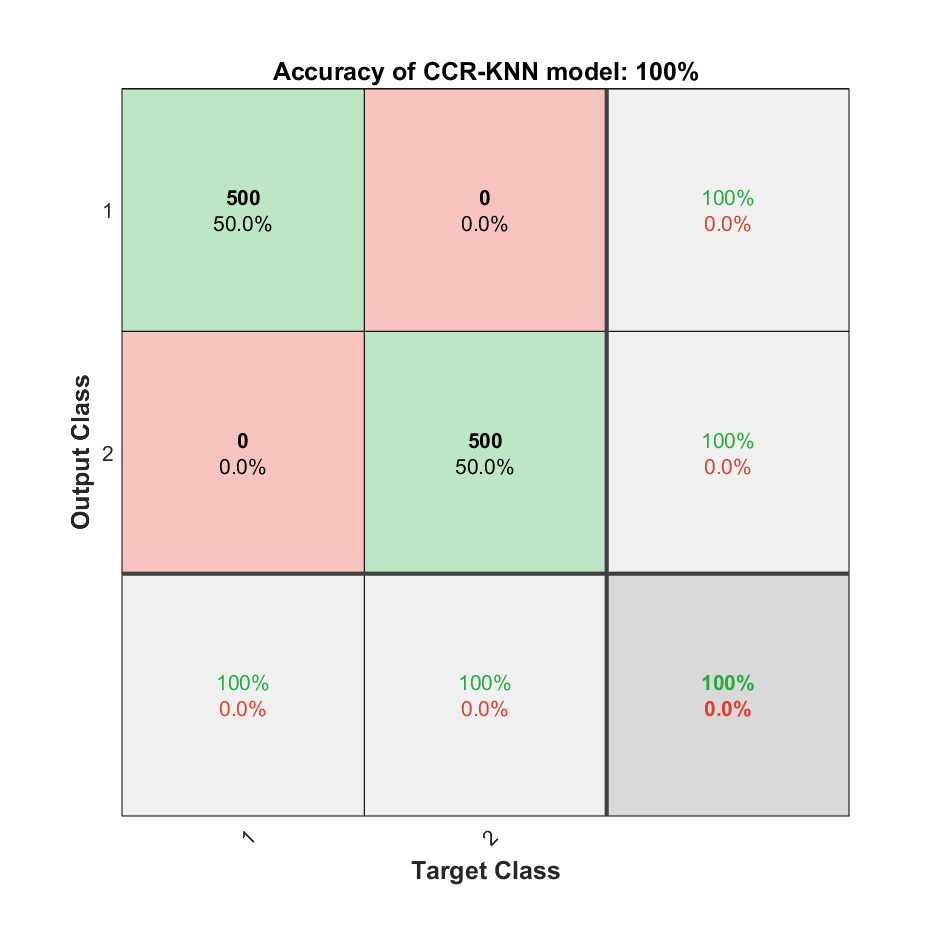}
    \label{Fig:Conf-CCR-KNN}
    }
    \subfigure[IOBCC-KNN]{
    \includegraphics[trim={1.2cm 1cm 2cm 2cm},clip,width=0.3\columnwidth]{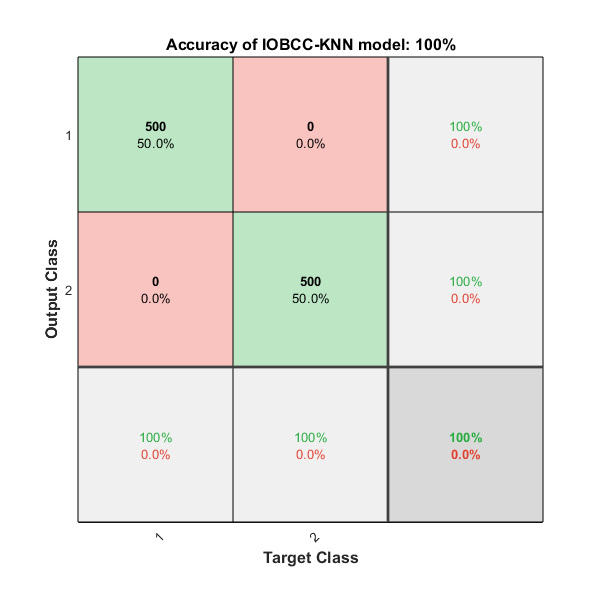}
    \label{Fig:Conf-IOBCC-KNN}
    }
    \subfigure[OOBCC-KNN]{
    \includegraphics[trim={1.2cm 1cm 2cm 2cm},clip,width=0.3\columnwidth]{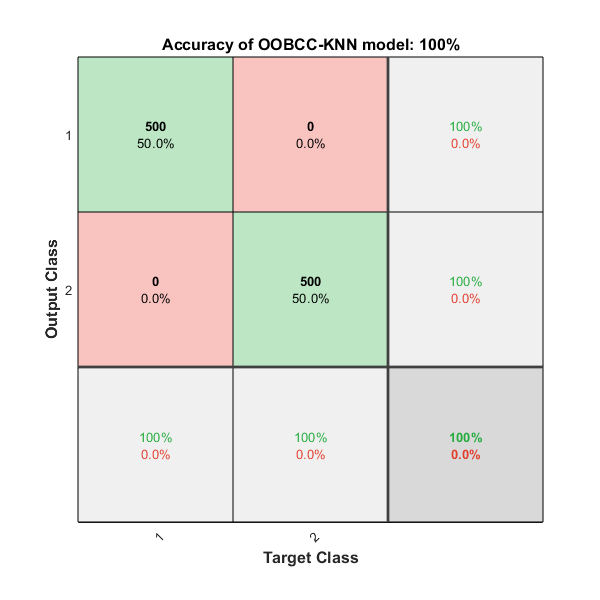}
    \label{Fig:Conf-OOBCC-KNN}
    }
    \subfigure[Additive-KNN]{
    \includegraphics[trim={1.2cm 1cm 2cm 2cm},clip,width=0.3\columnwidth]{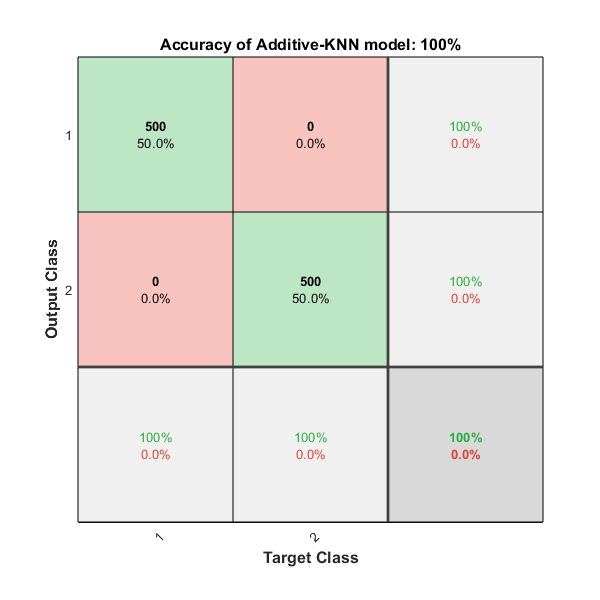}
    \label{Fig:Conf-Additive-KNN}
    }
    \caption{Confusion matrixes for different comparative KNN models.}
    \label{Fig:ConfusionResultsKNN}
    \end{center}
\end{figure}

\begin{figure}[ht!]
    \begin{center}
    \subfigure[CCR-SVM]{
    \includegraphics[trim={1.2cm 1cm 2cm 2cm},clip,width=0.3\columnwidth]{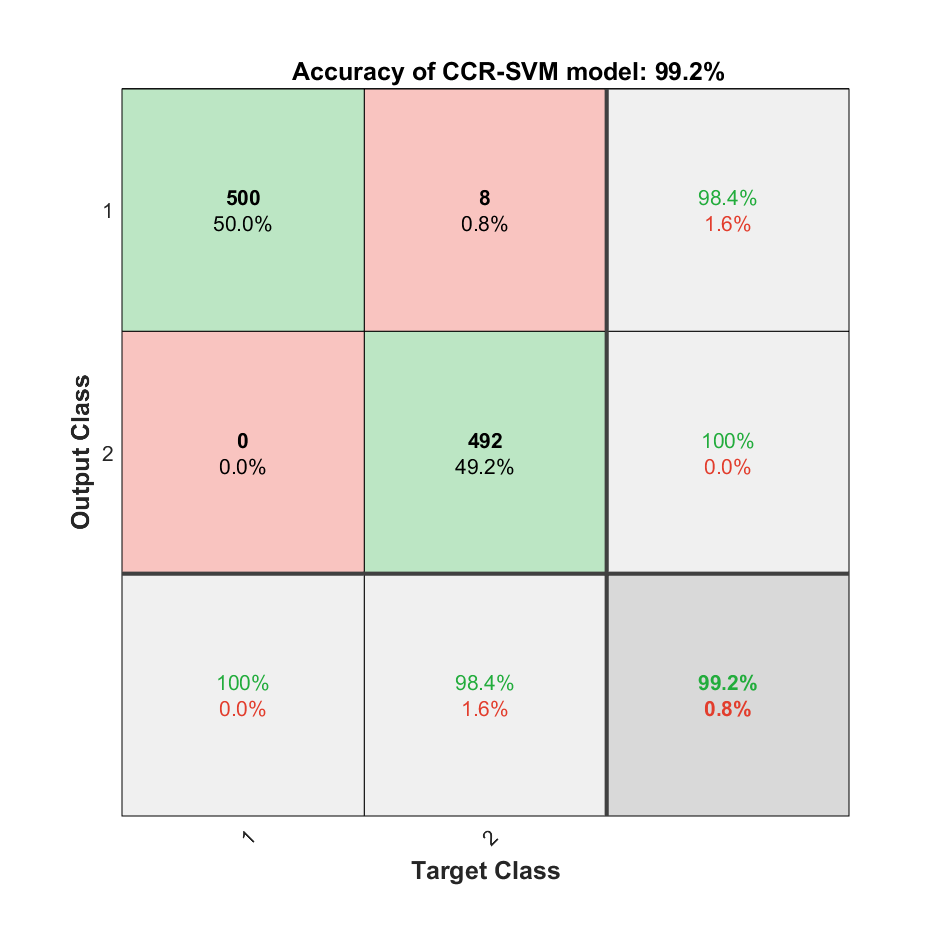}
    \label{Fig:Conf-CCR-SVM}
    }
    \subfigure[IOBCC-SVM]{
    \includegraphics[trim={1.2cm 1cm 2cm 2cm},clip,width=0.3\columnwidth]{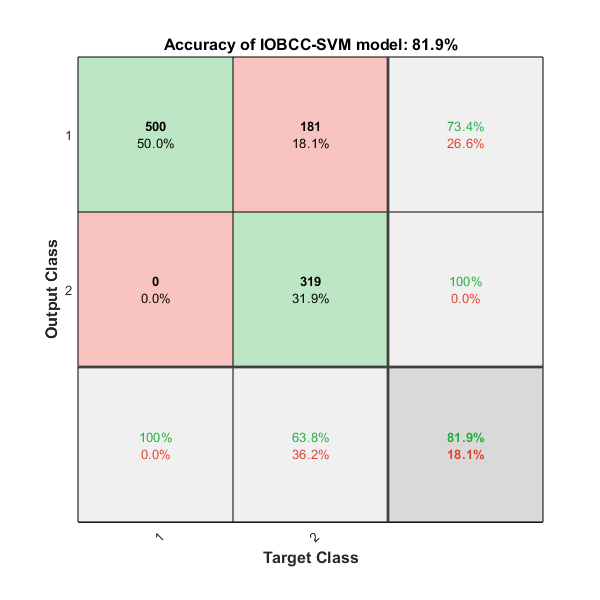}
    \label{Fig:Conf-IOBCC-SVM}
    }
    \subfigure[OOBCC-SVM]{
    \includegraphics[trim={1.2cm 1cm 2cm 2cm},clip,width=0.3\columnwidth]{Figures/Confusion_OOBCC_KNN.png}
    \label{Fig:Conf-OOBCC-SVM}
    }
    \subfigure[Additive-SVM]{
    \includegraphics[trim={1.2cm 1cm 2cm 2cm},clip,width=0.3\columnwidth]{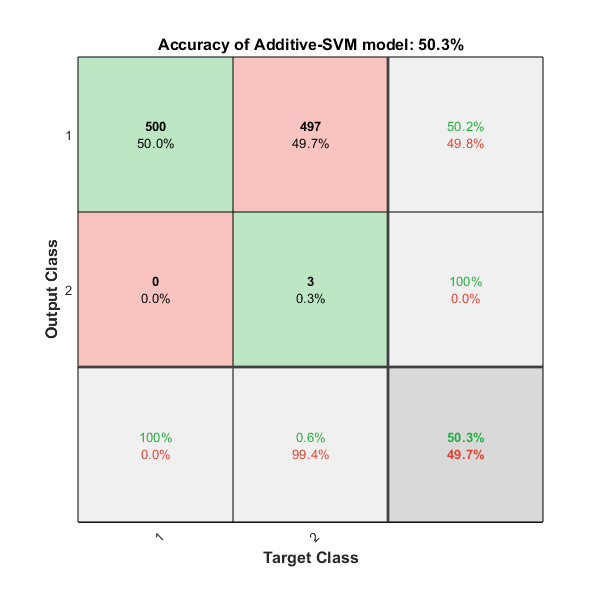}
    \label{Fig:Conf-Additive-SVM}
    }
    \caption{Confusion matrixes for different comparative SVM models.}
    \label{Fig:ConfusionResultsSVM}
    \end{center}
\end{figure}

\begin{figure}[ht!]
    \begin{center}
    \subfigure[CCR-naive Bayes]{
    \includegraphics[trim={1.2cm 1cm 2cm 2cm},clip,width=0.3\columnwidth]{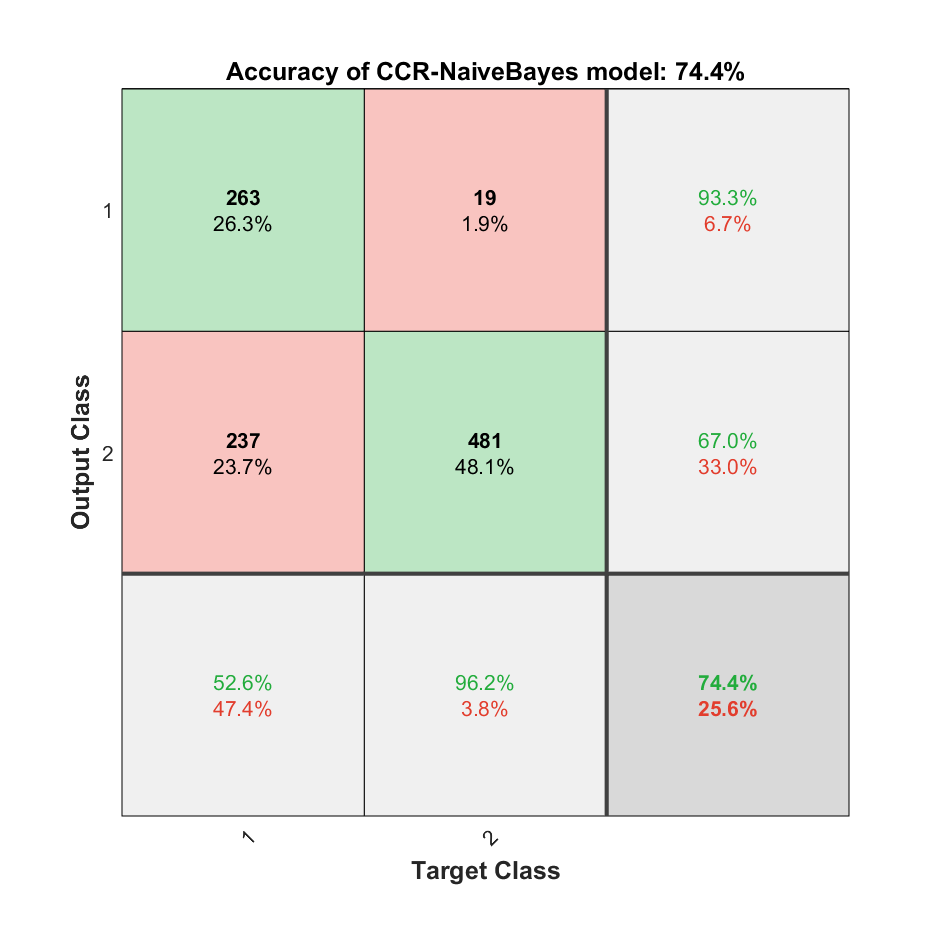}
    \label{Fig:Conf-CCR-NaiveBayes}
    }
    \subfigure[IOBCC-naive Bayes]{
    \includegraphics[trim={1.2cm 1cm 2cm 2cm},clip,width=0.3\columnwidth]{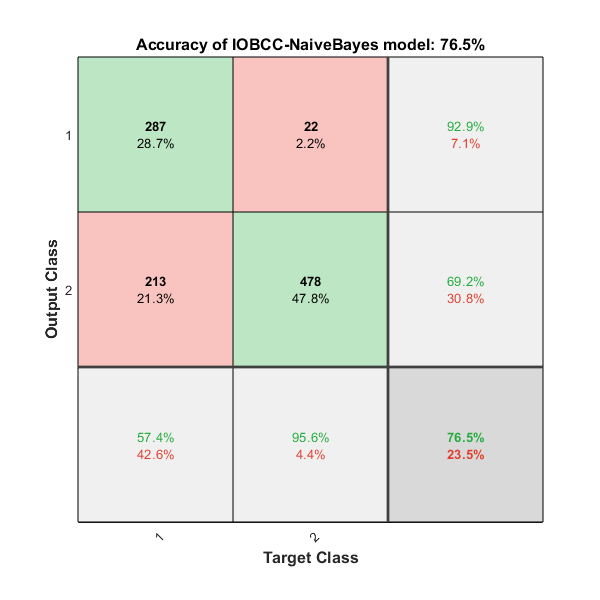}
    \label{Fig:Conf-IOBCC-NaiveBayes}
    }
    \subfigure[OOBCC-naive Bayes]{
    \includegraphics[trim={1.2cm 1cm 2cm 2cm},clip,width=0.3\columnwidth]{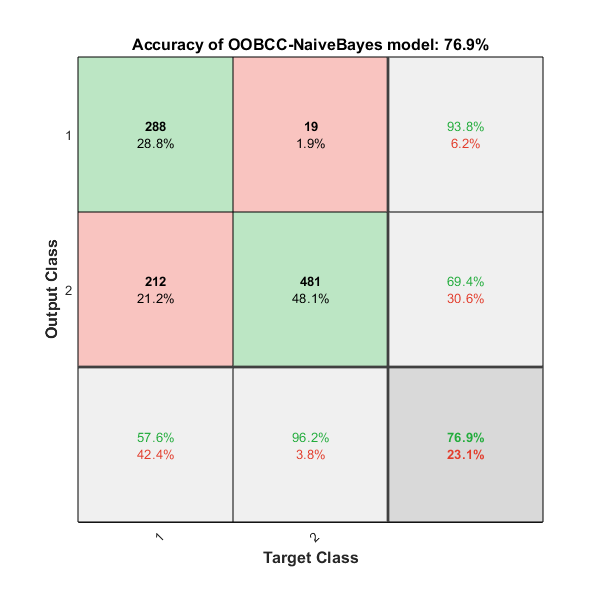}
    \label{Fig:Conf-OOBCC-NaiveBayes}
    }
    \subfigure[Additive-naive Bayes]{
    \includegraphics[trim={1.2cm 1cm 2cm 2cm},clip,width=0.3\columnwidth]{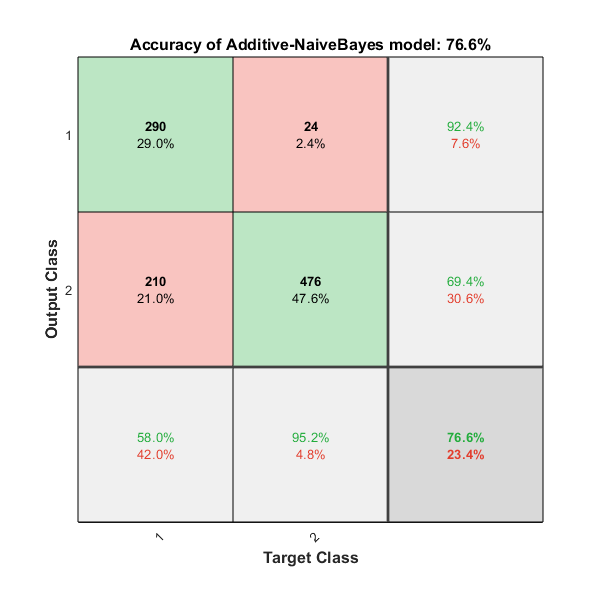}
    \label{Fig:Conf-Additive-NaiveBayes}
    }
    \caption{Confusion matrixes for different comparative naive Bayes models.}
    \label{Fig:ConfusionResultsNaiveBayes}
    \end{center}
\end{figure}

Table \ref{tab:ComparionDEAResults} compares the results obtained by SVM, KNN, and naive Bayes algorithms in comparison with the proposed sensor selection methods. Based on the classification results of various comparison methods, Table \ref{tab:ComparionDEAResults} summarizes the classification results for the healthy condition $(+)$ and the broken tooth condition $(-)$. It is evident from Table \ref{tab:ComparionDEAResults} that the CCR-KNN method performs better than others since the least number of sensors is used. A minimum number of sensors is selected using the CCR model(26 sensors), whereas a higher number is selected using the other DEA models(29 sensors). The CCR-KNN, the IOBCC-KNN, the OOBCC-KNN, the additive-KNN, and the OOBCC-KNN have better classification accuracy than other methods.
Figures \ref{Fig:ConfusionResultsKNN},\ref{Fig:ConfusionResultsSVM}, and \ref{Fig:ConfusionResultsNaiveBayes} illustrate how the confusion matrix is used to evaluate the classifier's performance. As a result of the confusion matrix, it is possible to accurately predict whether a healthy for the additive-SVM, the OOBCC-SVM, the IOBCC-SVM, the CCR-SVM, the OOBCC-KNN, the additive-KNN, the IOBCC-KNN, and CCR-KNN methods. Further, the CCR-KNN, the IOBCC-KNN, the OOBCC-KNN, the additive-KNN, and the OOBCC-KNN are the most accurate methods for identifying broken tooth conditions.
\begin{table}[ht!]
\tiny{
    \centering
     \caption{The comparison results of DEA models.}
    \label{tab:ComparionDEAResults}
    \begin{tabular}{cccccccccc}
    \hline
        Method &\# selected sensors&Accuracy(\%)&$Recall^{+}$&$Recall^{-}$&$Precision^{+}$&$Precision^{-}$&$F-score^{+}$&$F-score^{-}$&AUC\\
        \hline
        \textcolor{red}{CCR-KNN}&\textcolor{red}{26} &\textcolor{red}{100} &\textcolor{red}{1} &\textcolor{red}{1} &\textcolor{red}{1}& \textcolor{red}{1}& \textcolor{red}{1}&\textcolor{red}{1} &\textcolor{red}{1}\\
        IOBCC-KNN&29 &100 &1 &1 &1& 1& 1&1 &1\\
        OOBCC-KNN&29 &100 &1 &1 &1& 1& 1&1 &1\\
        Additive-KNN&29 &100 &1 &1 &1& 1& 1&1 &1\\
        CCR-SVM&26 &99.2 &1 &0.984 &0.9843& 1& 0.9921&0.9919 &0.992\\
        IOBCC-SVM&29 &81.9 &1 &0.638 &0.7342& 1& 0.8467&0.779 &0.819\\
        OOBCC-SVM&29 &100 &1 &1 &1& 1& 1&1 &1\\
        Additive-SVM&29 &50.3 &1 &0.006 &0.5015& 1& 0.668&0.0119 &0.503\\
        CCR-NaiveBayes&26 &74.4 &0.526 &0.962 &0.9326& 0.6699& 0.6726&0.7898 &0.744\\
        IOBCC-NaiveBayes&29 &76.5 &0.574 &0.956 &0.9288& 0.6918& 0.7095&0.8059 &0.765\\
        OOBCC-NaiveBayes&29 &76.9 &0.576 &0.962 &0.9381& 0.6941& 0.7138&0.8064 &0.769\\
        Additive-NaiveBayes&29 &76.6 &0.58 &0.952 &0.9236& 0.6939& 0.7125&0.8027 &0.766\\
       \hline
    \end{tabular}
    }
\end{table}

\subsection{Discussions}\label{subsec:Discussion}
\begin{figure}[ht!]
    \begin{center}
    \subfigure[Pearson-KNN]{
    \includegraphics[trim={1.2cm 1cm 2cm 2cm},clip,width=0.29\columnwidth]{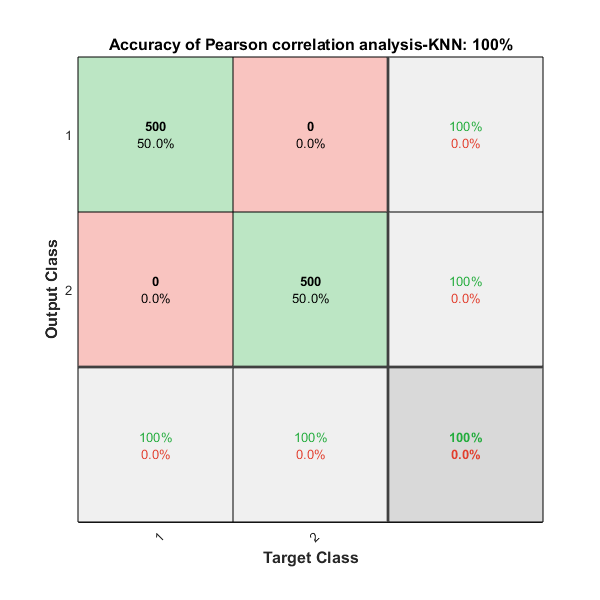}
    \label{Fig:Conf-Pearson-KNN}
    }
    \subfigure[Pearson-SVM]{
    \includegraphics[trim={1.2cm 1cm 2cm 2cm},clip,width=0.29\columnwidth]{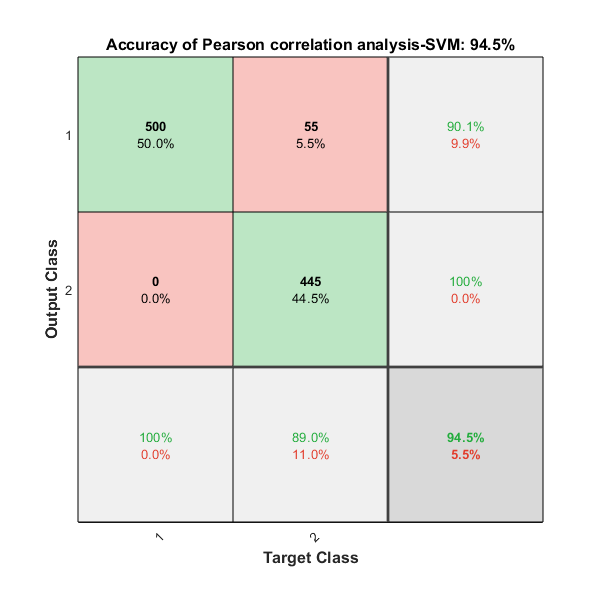}
    \label{Fig:Conf-Pearson-SVM}
    }
    \subfigure[Pearson-naive Bayes]{
    \includegraphics[trim={1.2cm 1cm 2cm 2cm},clip,width=0.3\columnwidth]{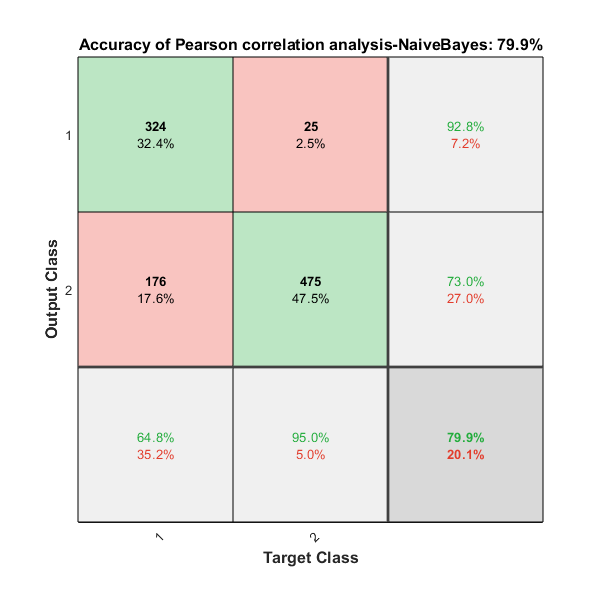}
    \label{Fig:Conf-Pearson-NaiveBayes}
    }
    \caption{Confusion matrixes for different comparative Pearson methods.}
    \label{Fig:ConfusionResultsPearson}
    \end{center}
\end{figure}

\begin{figure}[ht!]
    \begin{center}
    \subfigure[ETC-KNN]{
    \includegraphics[trim={1.2cm 1cm 2cm 2cm},clip,width=0.29\columnwidth]{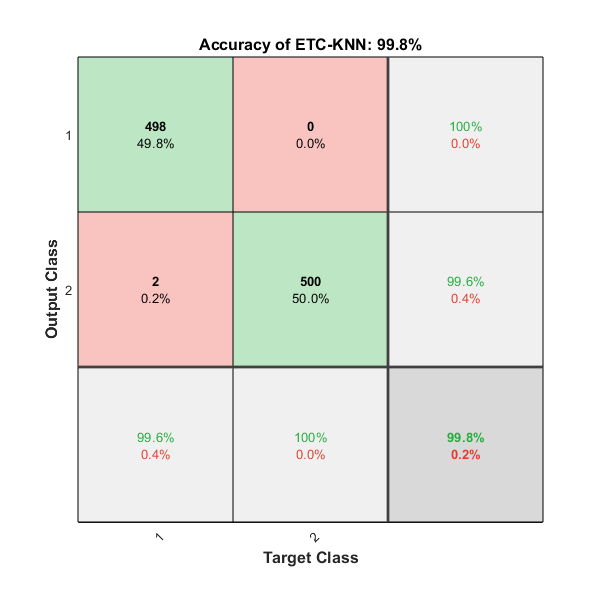}
    \label{Fig:Conf-ETC-KNN}
    }
    \subfigure[ETC-SVM]{
    \includegraphics[trim={1.2cm 1cm 2cm 2cm},clip,width=0.29\columnwidth]{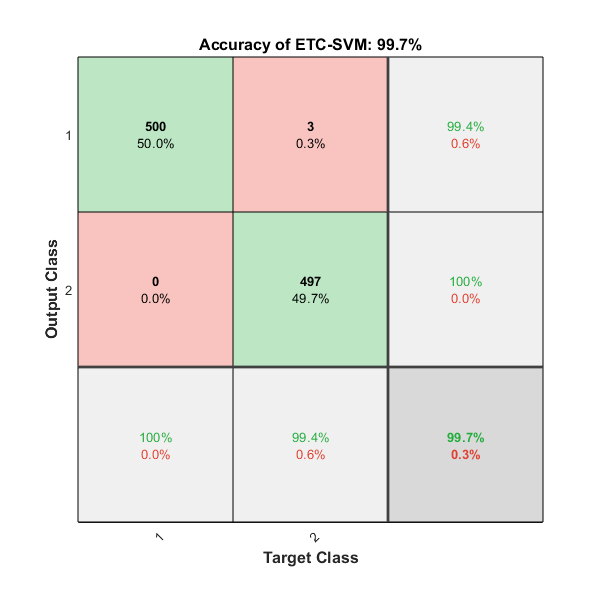}
    \label{Fig:Conf-ETC-SVM}
    }
    \subfigure[ETC-naive Bayes]{
    \includegraphics[trim={1.2cm 1cm 2cm 2cm},clip,width=0.29\columnwidth]{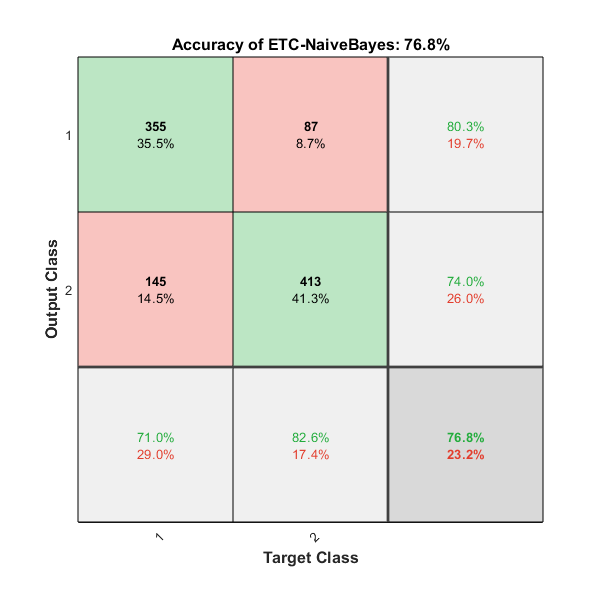}
    \label{Fig:Conf-ETC-NaiveBayes}
    }
    \caption{Confusion matrixes for different comparative ETC methods.}
    \label{Fig:ConfusionResultsETC}
    \end{center}
\end{figure}

\begin{figure}[ht!]
    \begin{center}
    \subfigure[mRMR-KNN]{
    \includegraphics[trim={1.2cm 1cm 2cm 2cm},clip,width=0.29\columnwidth]{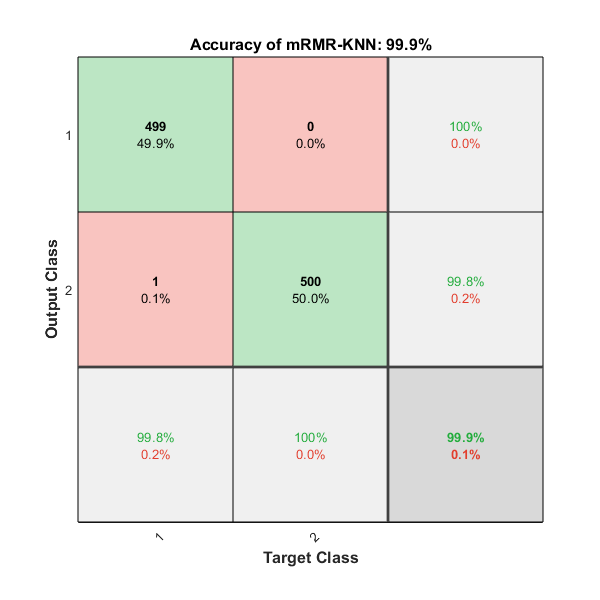}
    \label{Fig:Conf-mRMR-KNN}
    }
    \subfigure[mRMR-SVM]{
    \includegraphics[trim={1.2cm 1cm 2cm 2cm},clip,width=0.29\columnwidth]{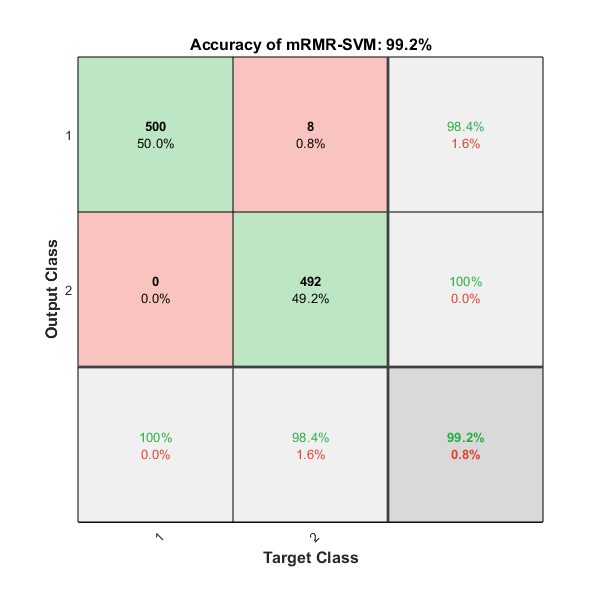}
    \label{Fig:Conf-mRMR-SVM}
    }
    \subfigure[mRMR-naive Bayes]{
    \includegraphics[trim={1.2cm 1cm 2cm 2cm},clip,width=0.29\columnwidth]{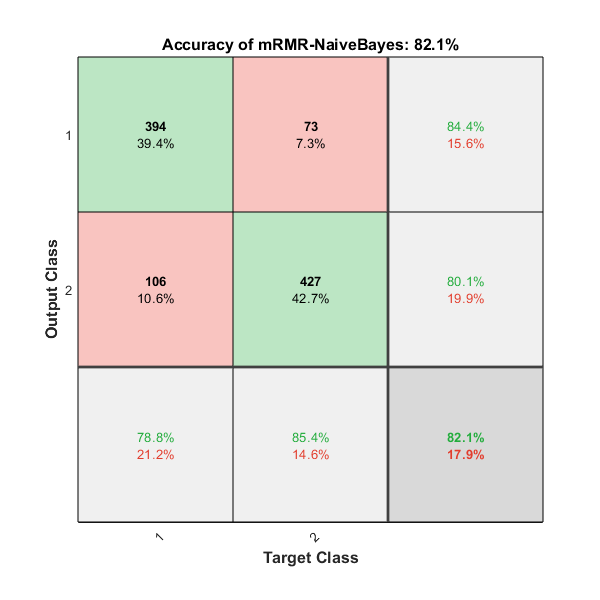}
    \label{Fig:Conf-mRMR-NaiveBayes}
    }
    \caption{Confusion matrixes for different comparative mRMR methods.}
    \label{Fig:ConfusionResultsmRMR}
    \end{center}
\end{figure}

We compare the performance of the CCR-KNN method with others, including the correlation analysis method(Pearson), the mRMR, and the ETC algorithms with SVM, KNN, and the naive Bayes classifiers. The performance results are summarized in Table \ref{tab:ComparionResults}. Despite selecting fewer sensors than the CCR method, the ETC algorithm produces lower classification performance than our proposed method. Pearson-KNNs exhibit similar performance results to CCR-KNN; however, it involves utilizing the maximum number of sensors, resulting in overfitting, complexity, and lower generalizability in the learning process.
Figures \ref{Fig:ConfusionResultsPearson},\ref{Fig:ConfusionResultsETC}, and \ref{Fig:ConfusionResultsmRMR} illustrate the possibility of accurately determining whether a system is healthy with the Pearson-KNN, the Pearson-SVM, and the ETC-SVM. Additionally, only the Person-KNN and the ETC-KNN provide the most accurate method of identifying broken tooth states.
\begin{table}[ht!]
\tiny{
    \centering
     \caption{The results of the comparison between the proposed and other methods.}
    \label{tab:ComparionResults}
    \begin{tabular}{cccccccccc}
    \hline
        Method &\# selected sensors&Accuracy(\%)&$Recall^{+}$&$Recall^{-}$&$Precision^{+}$&$Precision^{-}$&$F-score^{+}$&$F-score^{-}$&AUC\\
        \hline
        \textcolor{red}{CCR-KNN}&\textcolor{red}{26} &\textcolor{red}{100} &\textcolor{red}{1} &\textcolor{red}{1} &\textcolor{red}{1}& \textcolor{red}{1}& \textcolor{red}{1}&\textcolor{red}{1} &\textcolor{red}{1}\\
        Pearson-KNN&40 &100 &1 &1 &1& 1& 1&1 &1\\
        Pearson-SVM&40 &93.8 &1 &0.876 &0.8897& 1& 0.9416&0.9288 &0.938\\
        Pearson-NaiveBayes&40 &80.1 &0.656 &0.054 &0.9239& 0.0761& 0.7672&0.6317 &0.801\\
        ETC-KNN&16 &99.8 &0.996 &1 &1& 0.996& 0.998&0.998 &0.998\\
        ETC-SVM&16 &99.7 &1 &0.994 &0.994& 1& 0.997&0.997 &0.997\\
        ETC-NaiveBayes&16 &76.1 &0.702 &0.82 &0.7959& 0.7335& 0.746&0.7743 &0.761\\
        mRMR-KNN&21 &99.9 &0.998 &1 &1& 0.998& 0.999&0.999 &0.999\\
        mRMR-SVM&21 &99.6 &1 &0.992 &0.9921& 1& 0.996&0.996 &0.996\\
        mRMR-NaiveBayes&21 &81.4 &0.774 &0.854 &0.8413& 0.7907& 0.8062&0.8211 &0.814\\
       \hline
    \end{tabular}
    }
\end{table}

\section{Conclusions and prospects}\label{sec:Conclusion}
In this article, a DEA-based sensor selection framework is presented for the design of a fault diagnostic system. The healthy and faulty states are classified using supervised machine learning classifiers (SVM, KNN, and naive Bayes). This paper provides a case study demonstrating the proposed approach's effectiveness based on the gearbox's vibration signal. As a result of the computational results, it has been indicated that  DEA-based sensor selection approaches result in fewer sensors and a greater degree of accuracy in identifying faults. As a result of this study, the DEA-based sensor selection approaches are applicable in other case studies. For future research, this may serve as an approach for the efficient design of sensor selection for fault diagnosis.
In addition, it is an important future direction to include the dynamics in modelling, estimation, and sensor selection. Feedback control should consider this. A future extension of this research will focus on metrics relating to observability and controllability.
\subsection*{Acknowledgements}
The authors have declared no conflicts of interest regarding their research, authorship, or publication of
this article.
\typeout{}
\bibliographystyle{abbrv}
\bibliography{Refs}
\end{document}